\documentstyle[11pt]{article}

\newcommand{\be}{\begin{equation}}
\newcommand{\ee}{\end{equation}}
\newcommand{\bea}{\begin{array}}
\newcommand{\ea}{\end{array}}
\newcommand{\beqa}{\begin{eqnarray}}
\newcommand{\eeqa}{\end{eqnarray}}
\newcommand{\bean}{\begin{eqnarray*}}
\newcommand{\eean}{\end{eqnarray*}}
\newcommand{\eqn}[1]{(\ref{#1})}

\newcommand{\del}{\partial}
\newcommand{\nn}{\nonumber}

\def\up#1{\leavevmode \raise.16ex\hbox{#1}}

\newcommand{\gapproxeq}{\lower .7ex\hbox{$\;\stackrel{\textstyle >}{\sim}\;$}}
\newcommand{\lapproxeq}{\lower .7ex\hbox{$\;\stackrel{\textstyle <}{\sim}\;$}}

\newcounter{appendice}

\def\thebibliography#1{{\bf REFERENCES\markboth
 {REFERENCES}{REFERENCES}}\list
 {[\arabic{enumi}]}{\settowidth\labelwidth{[#1]}\leftmargin\labelwidth
 \advance\leftmargin\labelsep
 \usecounter{enumi}}
 \def\newblock{\hskip .11em plus .33em minus -.07em}
 \sloppy
 \sfcode`\.=1000\relax}

\begin{document}

\title{\hfill $\mbox{\small{
$\stackrel{\rm\textstyle hep-th/9806164}
{\rm \textstyle DSF-19/98}$}}$\\[1truecm]
Probability Representation in Quantum Field Theory}
\author{ 
V. I. Man'ko,\thanks{On leave from Lebedev Physical Institute, Moscow, Russia
}\,
L. Rosa, and P. Vitale\thanks{manko,rosa,vitale@na.infn.it}\\
%
%
  Dipartimento di Fisica dell'Universit\`a, and INFN Sez. di Napoli,\\
Mostra d'Oltremare, Pad.19, I-80125, Napoli, Italy }
\maketitle
\begin{abstract}
The recently proposed probability representation of quantum mechanics
is generalized to  quantum field theory. We introduce a probability
distribution functional for field configurations and find an evolution
equation for such a distribution. The connection to the time-dependent
generating functional of Green's functions is elucidated and the
classical limit is discussed. 
\end{abstract}

The use of statistical methods for describing quantum physics gives the
opportunity to describe classical and quantum phenomena in a unified
approach. Since the beginning of quantum mechanics there have been
attempts to understand its nature in a classical-like context, namely
to describe quantum states in terms of a classical distribution of
probability. It is this philosophy which inspired the so-called
quasi-probability distribution functions of Wigner, Husimi, Glauber and
Sudarshan \cite{wig32}. The original goal was not completely
achieved (the above distribution functions are not always positive
defined or they do not describe measurable variables) until Cahill and
Glauber in \cite{cahill} introduced a class of distribution functions,
known as marginal distribution functions (MDF), which enjoyed all the
properties of a density of probability.  Nevertheless, it was realized
only recently \cite{tom3} that quantum mechanics could be described
entirely in terms of  a distribution of such a family, suitably defined
for a random variable, which we  will specify below. In \cite{tom3} 
a consistent
scheme has been proposed, the so-called probability representation, 
which has been shown to be completely equivalent to the
ordinary formulation. Quantum states are described by a distribution of
probability, the MDF, and
the time evolution by an integro-differential equation for the MDF.
Invertible relations have been established between the MDF and the
density matrix \cite{tom3,tom1}  and between the Green's functions of
the related evolution equations \cite{ovman}. 

In this framework classical and quantum phenomena, both statistically
described, only differ by the evolution equations of the distributions 
of probabilities for the relevant observables. The quantum evolution
equation, of Fokker-Plank type, is seen to reduce to Boltzmann equation
for the classical distribution of probability when the classical limit
is considered. 

In this letter we generalize the probability representation first to
the case of $N$ interacting particles, then to non relativistic quantum
field theory. We consider a system of interacting oscillators which
describe, when the limit to the continuum is performed, a
self-interacting scalar field theory with generic self-interaction
potential. We introduce the notion of MDF for the quantum state of such
system and derive the evolution equation both for the discrete and
continuous cases. The MDF is seen to be a distribution of probability
with the same arguments used for 1-d quantum mechanics. 

Interestingly similar ideas have been developed by Wetterich in
\cite{wet2} in connection with the approach to equilibrium in
non-equilibrium quantum field theories. There an evolution equation is
found for a suitably defined time-dependent generating functional
\cite{zinn}.  We establish the connection between our MDF and
Wetterich's generating functional in terms of a (functional-) Fourier
transform. 

In section {\bf 1} we briefly review the probability representation of
quantum mechanics for the simple case of a one-dimensional quadratic
Hamiltonian. The derivation of the evolution equation for the MDF is
performed in detail in a slightly different manner from its original
derivation \cite{tom3}, but more suitable for generalizations. In
section {\bf 2} we consider a quadratic Hamiltonian describing $N$
interacting particles. We define the MDF and show that this is a well
defined probability distribution. We then find the evolution equation.
In section {\bf 3} we consider a scalar field theory with
self-interacting potential, which may be seen as the limit to the
continuum  of the previous model. We define the MDF which is  now a
probability distribution functional, and derive its exact evolution
equation. 
In section {\bf 4} we derive the evolution equations for the above
mentioned systems, directly in terms of the Fourier transform of the
MDF, the quantum characteristic function. We show then how our results
may be connected to those found in \cite{wet2}. 

\section{The Probability Representation of Quantum Mechanics} 

The MDF of a random variable $X$ was introduced in \cite{cahill} as the
Fourier transform of the quantum  characteristic function  
$\chi (k) = <e^{ik\hat X}>$, to be  
\be 
w(X,t)= {1\over 2\pi} \int dk\, e^{-ikX}
<e^{ik\hat X}>~ , \label{mdf} 
\ee 
where $\hat X$ is the operator
associated to $X$, $<\hat A> = \mbox{\rm Tr} (\hat\rho \hat
A)$, and $\hat\rho$ is the time-dependent density operator. 
It is shown in \cite{cahill} 
that $w(X,t)$ is positive and normalized to unity, provided $\hat X$
is an observable. This theorem may be easily proven taking for 
simplicity $\hat\rho$ to be the density operator for a pure state. 
Then, evaluating the trace in \eqn{mdf} on eigenstates of the operator 
$\hat X$, it can be verified that \eqn{mdf} yields  $w(X,t)= \rho(X,X,t)$
which is positive and normalized to unity.

We recall that the quantum characteristic function is, up to factors
of $i$,  the generating function of  the momenta of any order, for the
probability distribution of the operator $\hat X$. Hence it plays in
quantum statistical mechanics the same r\^ole as the generating
functionals for the Green's functions in quantum field theory. In ref.
\cite{tom1} $X$ is taken to be a variable of the form 
\be
 X=\mu q + \nu p \, ,\label{x}
\ee
where $\mu, ~\nu$ are real parameters labelling different reference
frames in the phase space. $\mu$ is dimensionless, while $[\nu]=
[m^{-1}] [t] $.  Thus, $X$ represents the position coordinate taking
values in an ensemble of reference frames. For such a choice of $
X$ it was shown that there exists an invertible relation among the MDF
and the density matrix, respectively in \cite{tom1} for the 1-d case, 
and in \cite{dar} for the 2-d case. 
This relation was originally understood through the Wigner function: 
the MDF was
expressed in terms of the Wigner function which is in turn related to 
the density matrix and viceversa. The evolution equation of
the MDF was then found starting from an evolution equation for the
Wigner function established by Moyal in \cite{moyal}. This intermediate
step in terms of the Wigner function is  not necessary. We can directly
invert \eqn{mdf}, when the variable $X$ and the associated operator are
given by \eqn{x}. The evolution equation is then obtained (in the
Schr\"odinger representation) by means of the Liouville equation for
the density operator, in coordinate representation. In view of the
subsequent generalization to $N$ degrees of freedom and to field
theory, let us derive these results in some detail for a one
dimensional system. Equation \eqn{mdf} is explicitly written as 
\beqa
w(X,\mu,\nu,t)&=& {1\over 2\pi} \int dk\, \int dZ \, e^{-ikX}
<Z|\hat\rho e^{ik\hat X}|Z>~ \nn\\ 
&=& {1\over 2\pi} \int dk\, \int dZ \, \rho(Z, Z-k\nu\hbar) 
e^{-ik[X-\mu(Z-k\nu\hbar/2)]} .
\eeqa   
The MDF so defined is normalized with respect to the $X$ variable: $ 
\int dX w(X,\mu,\nu,t)=1$.
Performing the change of variables $~Z'=Z,~Z''=Z-k\nu\hbar$ we may reexpress 
the MDF in the more convenient form:
\be
w(X,\mu,\nu,t)={1\over 2\pi |\nu|\hbar} \int \rho(Z',Z'',t) \exp 
\left[-i{Z'-Z''\over \nu\hbar}\left(X-\mu {Z+Z'\over 2}
\right)\right] dZ'\, dZ'' \label{wro}
\ee
which can be inverted to 
\be
\rho (X,X',t)=|\alpha| \int w(Y,\mu,{X-X'\over \hbar\alpha}) \exp \left[ 
i\alpha\left(Y-\mu{X+X'\over 2}\right)\right] d\mu \, dY \label{row1}
\ee
where $\alpha$ is a parameter with dimension of an inverse length.
The density matrix is independent of $\alpha$. In facts, using the 
homogeneity of the MDF, $w(\alpha X, \alpha\mu,\alpha\nu)=|\alpha|^{-1} 
w(X, \mu,\nu)$, which is evident from the definition, 
\eqn{row1} may be written as
\be
\rho (X,X',t)=\int w(Y,\mu, X-X') \exp \left[ 
{i\over \hbar} \left(Y-\mu{X+X'\over 2}\right)\right] d\mu \, dY \label{row}
\ee
where the variables $Y,\mu$ have been rescaled by $\alpha$.
It is important to note that, for \eqn{wro} to be invertible, it is 
necessary that $X$ be a coordinate variable taking values in an 
ensemble of phase 
spaces; in other words, the specific choices $\mu=1, \nu=0$ or any 
other fixing of the parameters $\mu$ and $\nu$ 
would not allow to reconstruct the density matrix.  
Hence, the MDF contains the same amount of information on a quantum 
state as the density matrix, only if Eq. \eqn{x} is assumed. 

We now address the problem of finding the evolution equation for the 
MDF, for Hamiltonians of the form
\be
\hat H = {{\hat p}^2\over 2m} +V(\hat q).
\ee
Using the Liouville equation
\be
\frac{\del \hat\rho}{\del t} + {i\over \hbar} 
[\hat H,\hat\rho]=0 \label{liou}
\ee
and substituting into Eq. \eqn{wro}, we have 
\beqa
{\dot w}(X,\mu,\nu,t)&=&-{i\over 2\pi |\nu|\hbar} \int \left[ -{\hbar^2\over 
2m}\left(
\frac{\del^2}{\del Z^2} - \frac{\del^2}{\del Z'^2} \right) 
\left( V(Z)- V(Z')\right)\right] 
\rho(Z,Z',t) \nn \\
& &\times\exp \left[-i{Z-Z'\over 
\nu\hbar}\left(X-\mu{Z+Z'\over 2}\right)\right] dZ\, d Z' ~.
\eeqa
Integrating by parts and assuming the density matrix to be zero at 
infinity, we finally have
\beqa
{\dot w}(X,\mu,\nu,t) &=& \left\{{1\over m} 
\mu {\del\over \del \nu}+{i\over \hbar} \left[ 
V\left(-({\del \over \del X})^{-1} {\del \over \del \mu} 
-{i\nu\hbar\over 2} {\del\over 
\del X}\right) \right.\right.\nn\\ 
&-&\left.\left. V\left(-({\del \over \del X})^{-1} {\del \over \del \mu} 
+{i\nu\hbar\over 2} {\del\over \del X}\right)\right]\right\} w(X,\mu,\nu,t)~,
\label{evo}
\eeqa
where the operator $({\del \over \del X})^{-1}$ is so defined
\be
({\del \over \del X})^{-1}  \int f(Z)  e^{g(Z)X}dZ =
\int {f(Z)\over g(Z)}  e^{g(Z)X} dZ~. 
\label{invder}
\ee
This equation, which plays the r\^ole of the Schr\"odinger equation in 
the alternative scheme just outlined, has been studied and solved for 
some quantum 
mechanical systems \cite{noipra},\cite{manko}. 
The classical limit of \eqn{evo} is easily seen to be 
\be
{\dot w}(X,\mu,\nu,t)= \left\{\frac{\mu}{m} {\del\over \del \nu}+\nu 
V'\left(-\left({\del \over \del X}\right)^{-1} {\del \over \del 
\mu}\right){\del \over \del X} \right\} w(X,\mu,\nu,t)~,
\label{evocl}
 \ee
where $V'$ is the derivative of the potential with respect to the 
argument. Equation \eqn{evocl} 
may be checked to be equivalent to Boltzmann equation 
for a classical distribution of probability $f(q,p,t)$ ,
\be
{\del f\over \del t} + {p\over m} {\del f \over \del q} -  {\del V \over 
\del q} {\del f \over \del p} =0,
\ee
after performing the change of variables
\be
w(X,\mu,\nu,t) = {1\over 2\pi} \int f(q,p,t) e^{ik(X-\mu q -\nu p) } 
dk~ dq~ dp~;
\ee
Hence, the classical and quantum evolution equations only differ by 
terms of higher order in $\hbar$. Moreover, for potentials quadratic in 
$\hat q$, higher order terms cancel out and the quantum evolution 
equation coincides with the classical one. 
This leads to the remarkable result that there is no difference between 
the evolution of the distributions of probability for quantum and 
classical observables, when the system is described by a Hamiltonian 
quadratic in positions and momenta. For this kind of systems, the 
propagator is the same  \cite{ovman}. Of course, what makes the difference 
is the initial condition.

\section{Generalization to $N$ degrees of freedom}
We consider now a system of $N$ interacting particles sitting on the 
sites of a lattice (we choose it to be one-dimensional for simplicity).
The Hamiltonian of the system is 
\be
\hat H = \sum_{i=1}^N {{\hat p_i}^2\over 2m_i} + V(\hat q).\label{hamn}
\ee
We assume the masses to be equal to unity. We take the potential 
to be of the form:
\be
V(\hat q)= \sum_{i=1}^N \left ( {1\over{2a^2}} (\hat q_{i+1} - \hat q_i)^2
+ U(\hat q_i)\right)
\ee
where $a$ is the lattice spacing and $U(\hat q_i)$ is the part of the 
potential which depends only on the position of the i-th particle.
This specification is not essential for the purposes of this section, 
but it will become necessary for understanding the limit to the 
continuum, which will be considered in next section.
The quantum  characteristic function  for the $N$ dimensional system 
may be defined as 
\be
\chi (k_1,...k_N) = <e^{i\sum_{i} k_i\hat X_i}>~,\label{chin}
\ee
where $<\hat A> = \mbox{\rm Tr} (\hat\rho \hat
A)$, and $\hat\rho$ is the density operator of the system. (In case 
there is no interaction between different sites of the lattice the 
density operator may be factorized and the characteristic function is 
just the product $ \chi (k_1,...k_N) = \Pi_{i=1}^N \chi(k_i) $ .) 

Performing the Fourier transform of \eqn{chin} the MDF is 
then given by
\be 
w(X_\sigma,\mu_\sigma,\nu_\sigma,t)= {1\over (2\pi)^N} \int dk_1 ... dk_N\, 
e^{-i\sum_ik_iX_i}
<e^{i\sum_ik_i\hat X_i}>~ ; \label{mdfn} 
\ee 
where $\sigma$ is a collective index. It may be shown 
that this is a probability distribution, namely that it is positive 
definite and normalized, provided $\hat X_i$ are observables. The proof 
goes along with the one-dimensional case. We first suppose that there 
is no interaction between different sites at some initial time $t_0$ 
and we assume for simplicity that the system be in a pure state $|\psi> 
= |\psi_1>\otimes ... \otimes |\psi_N>$. Using the factorization 
property of the quantum characteristic function Eq. \eqn{mdfn} may be 
seen to reduce to the product $w(X_\sigma,\mu_\sigma,\nu_\sigma,t_0)= 
\prod_i\rho_i(X_i,X_i,t_0)=\rho(X,X,t_0)$, which is positive and 
normalized. Then Liouville equation guarantees that this result stays 
valid when the interaction is switched on.
In analogy with the one-dimensional case we now introduce the variables
\be
X_i = \mu_i q_i + \nu_i p_i \label{xi}
\ee
with $\hat X_i$ accordingly defined. 
 
Introducing the notation $|Z_\sigma>\equiv |Z_1>\otimes...\otimes|Z_N>$
we rewrite \eqn{mdfn} as
\beqa
&& w(X_\sigma,\mu_\sigma,\nu_\sigma,t)= {1\over (2\pi)^N} \int 
\prod_{i=1}^N dk_i 
\, dZ_i \, e^{-i\sum_{j=1}^N k_j X_j}
<Z_\sigma | \hat\rho e^{i \sum_{l=1}^N k_l\hat X_l}|Z_\sigma>~ \cr 
&&= {1\over (2\pi)^N} \int \Pi_{i=1}^N dk_i\, dZ_i \, \rho(Z_\sigma, 
Z_\sigma -k_\sigma \nu_\sigma\hbar ) 
e^{-i\sum_{j+1}^N k_j[X_j-\mu_j(Z_j-k_j\nu_j\hbar/2)]} ,
\eeqa   
where $\rho(Z_\sigma, Z'_\sigma)=  <Z_\sigma|\hat\rho |Z'_\sigma>$.
Performing the change of variables 
\be
Z'_i=Z_i~,~~ Z''_i= Z_i - k_i\nu_i\hbar
\ee
we have
\beqa
w(X_\sigma,\mu_\sigma,\nu_\sigma,t)&=&{1\over (2\pi\hbar )^N} \int 
\prod_{i=1}^N \left ({dZ'_i\, d Z''_i\over |\nu_i|}\right) 
\rho(Z'_\sigma,Z''_\sigma,t) \nn \\
&\times&\exp \left[-i\sum_{j=1}^N {Z'_j-Z''_j\over 
\nu_j\hbar}\left(X_j-\mu_j{Z'_j+Z''_j\over 2}\right)\right] \label{wron}
\eeqa
which can be inverted to 
\be
\rho (X_\sigma,X'_\sigma,t)={1\over (2\pi)^N} \int \prod_{i=1}^N d\mu_i \, dY_i
~w(Y_\sigma,\mu_\sigma,X_\sigma-X'_\sigma)
  \exp \left[ 
{i\over \hbar}\sum_{j=1}^N \left( Y_j-\mu_j{X_j+X'_j\over 2}\right)\right] . 
\label{rown}
\ee
Once again, we recall that the inversion of \eqn{wron} is made possible by
choosing the variables $X_i$ as in \eqn{xi}.
We now use the Liouville equation \eqn{liou} to get 
\beqa
{\dot w}(X_\sigma,\mu_\sigma,\nu_\sigma,t) &=& {-i\over (2\pi\hbar)^N} 
\int \prod_{i=1}^N\left({1\over |\nu_i| } dZ_i\, d Z'_i \right)
\Biggl[ 
\sum_{j=1}^N -{\hbar^2\over 2m}
\left( 
\frac{\del^2}{\del Z_j^2} - \frac{\del^2}{\del Z_j^{'2}}  
\right)  \\
&+&  \left(V(Z_\sigma) - V(Z'_\sigma)\right)\Biggr] 
\rho(Z_\sigma,Z'_\sigma,t) 
\exp \left[-i\sum_{l=1}^N {Z_l-Z'_l\over 
\nu_l\hbar}\left(X_l-\mu_l{Z_l+Z'_l\over 2}\right)\right] ~.\nn
\eeqa
Integrating by parts and assuming the density matrix to be zero at 
infinity, we finally have
\beqa
{\dot w}(X_\sigma,\mu_\sigma,\nu_\sigma,t) &=& \Biggl\{ 
\sum_{i=1}^N\mu_i {\del\over \del \nu_i}+{i\over \hbar} 
\left[ 
V\left(
-({\del \over \del X_\sigma})^{-1} {\del \over \del \mu_\sigma} 
-{i\nu_\sigma\hbar\over 2} {\del\over \del X_\sigma} 
\right) 
\right. \nn\\
&-& \left. 
V\left(
-({\del \over \del X_\sigma})^{-1} {\del \over \del \mu_\sigma} 
+{i\nu_\sigma\hbar\over 2} {\del\over \del X_\sigma}
\right)
\right] \Biggr\} w(X_\sigma,\mu_\sigma,\nu_\sigma,t)~,
\label{evon}
\eeqa
where the inverse derivative is defined as in \eqn{invder}.
We report for future convenience the term containing the potential when 
explicitating the interaction between neighbours:
\beqa
&&\left[ 
V\left(-({\del \over \del X_\sigma})^{-1} {\del \over \del \mu_\sigma} 
-{i\nu\hbar\over 2} {\del\over 
\del X_\sigma}\right) - V\left(-({\del \over \del X_\sigma})^{-1} 
{\del \over \del \mu_\sigma} 
+{i\nu_\sigma\hbar\over 2} {\del\over \del X_\sigma}\right)\right] \nn\\
&&= \left[ 
U\left(-({\del \over \del X_\sigma})^{-1} {\del \over \del \mu_\sigma} 
-{i\nu\hbar\over 2} {\del\over 
\del X_\sigma}\right) - U\left(-({\del \over \del X_\sigma})^{-1} 
{\del \over \del \mu_\sigma} 
+{i\nu_\sigma\hbar \over 2} {\del\over \del X_\sigma}\right)\right]\nn\\
&&- {i\hbar\over a^2} \sum_{i=1}^N \nu_i \left[ {\del \over \del X_i} 
\left({\del \over \del X_{i+1}}\right)^{-1} {\del \over \del \mu_{i+1}}
-2 {\del \over \del \mu_i} + {\del \over \del X_i} 
\left({\del \over \del X_{i-1}}\right)^{-1} 
{\del \over \del \mu_{i-1}}\right]~. \label{potenital}
\eeqa
When considering the classical limit we have
\beqa
{\dot w}(X_\sigma,\mu_\sigma,\nu_\sigma,t)&=& \left\{\sum_{i=1}^N\mu_i 
{\del\over \del \nu_i}+ \nu_i
{1\over a^2} \left[ {\del \over \del X_i} 
\left({\del \over \del X_{i+1}}\right)^{-1} {\del \over \del \mu_{i+1}}
\right.\right. \nn\\
&-& \left. \left. 2 {\del \over \del \mu_i} + {\del \over \del X_i} 
\left({\del \over \del X_{i-1}}\right)^{-1} 
{\del \over \del \mu_{i-1}}\right] \right. \nn\\
& & \left. +V_i\left(-({\del \over \del X_\sigma})^{-1} {\del \over \del 
\mu_\sigma} \right) {\del\over  \del X_\sigma}  \right\}
w(X_\sigma,\mu_\sigma,\nu_\sigma,t)~,\label{evoncl}
\eeqa
where $U_i$ is the derivative of the self-interaction potential with 
respect to the $i-th$ variable. 
Equation \eqn{evoncl} may be seen to be equivalent to the Boltzmann 
equation as in the one-dimensional case. Moreover, Hamiltonians which 
are quadratic in positions and momenta yield the same evolution 
equations for classical and quantum probability distributions.

\section{Generalization to Field Theory}
We now consider a scalar quantum field theory described by the 
Hamiltonian 
\be
\hat H = \int d^d x~ \left[{1\over 2} \hat\pi^2(x) + {1\over 2} 
\sum_{b=1}^d(\del_b\hat\phi(x))^2 + U(\hat\phi(x))\right] ~.
\label{hamfi}
\ee
$d$ is the spatial dimension, while 
$U(\phi(x))$ is the  self-interacting potential, polynomial in the 
field $\hat\phi$. The Hamiltonian \eqn{hamfi} is easily seen to be 
obtained by the discrete Hamiltonian \eqn{hamn} by taking the limit to 
the continuum ($a\rightarrow 0$) with the following rules:
\beqa
a^{-d/2}\hat q_i \rightarrow \hat\phi(x)~~&&~~
a^{-d/2}\hat p_i \rightarrow \hat\pi(x)\nn\\
a^{d}\sum_i \rightarrow \int d^d x~~&&~~
a^{-(d/2+1)}(\hat q_{i+1} - \hat q_i) \rightarrow {\del \hat \phi(x)\over 
\del x_b}~.
\eeqa
In analogy with the discrete case we introduce the field
\be
\hat\Phi(x)= \mu(x)\hat\phi(x) + \nu(x)\hat\pi(x)
\ee
where $\mu(x)=\lim_{a\rightarrow 0} a^{-d/2} \mu_i$, and 
$\nu(x)=\lim_{a\rightarrow 0} a^{-d/2} \nu_i$ .  

The quantum characteristic functional, which now will play the r\^ole 
of generating functional for correlation functions of the fields, may 
be defined as
\be
\chi(k(x))= <e^{i\int d^d x~ k(x) \hat\Phi(x)}>=\mbox{\rm Tr} \left(\hat\rho(t)
e^{i\int d^d x~ k(x) \hat\Phi(x)}\right)~.
\ee
The functional Fourier transform of $\chi(k(x))$, what we will call the
marginal distribution functional (MD${\cal F}$), still defines a
probability distribution. This can be understood by recognizing that it
is the limit of the MDF for the discrete $N$-dimensional system
considered in the previous section: 
\be
w(\Phi(x), \mu(x), \nu(x), t) = \int {\cal D}k\; e^{-i\int k(x) \Phi(x) dx} 
\chi(k)=\lim_{a\rightarrow 0} w(X_\sigma,\mu_\sigma,\nu_\sigma,t)~,
\label{mdfc}
\ee
where $\prod_i {dk_i\over (2\pi)^N }\rightarrow \int {\cal D} k$.
Also, the density matrix functional may be defined as the limit of Eq. 
\eqn{rown} to be
\be
\rho(\Phi, \Phi',t)= \int {\cal D} \mu {\cal D} \Psi \;
w(\Psi,\mu,\Phi-\Phi') \exp\left\{ {i\over \hbar} \int dy 
\left[\Psi(y)-\mu(y)\left({\Phi(y)-\Phi'(y)\over 2}\right)\right] \right\}~.
\ee
Then, the evolution equation for the probability distribution 
functional is easily obtained by taking the limit of \eqn{evon}: 
\beqa
&&{\dot w}(\Phi(x), \mu(x), \nu(x), t) = 
\left\{  \int d^d x~ 
\left[ \mu(x) {\delta\over \delta \nu(x)} + 2 \nu(x){\delta\over \delta
\Phi(x)}\Delta 
\left[
\left(
{\delta\over \delta \Phi(x)}
\right)^{-1}
{\delta\over \delta \mu(x)}   
\right]
\right]
\right.\nn\\
      &&+ \left. {i\over \hbar} 
\left[ U
\left[
\left( {-\delta\over \delta \Phi(x)}
\right)^{-1} 
{\delta\over\delta \mu(x)} - {i \nu(x)\hbar\over 2} {\delta\over \delta \Phi(x)}
\right] 
\right.\right.\nn\\
&-& \left. \left. U
\left[
\left(  {-\delta\over \delta \Phi(x)}
\right)^{-1} {\delta\over
\delta \mu(x)} + {i \nu(x)\hbar\over 2} 
{\delta\over \delta \Phi(x)}
\right] 
\right] 
\right\}
w(\Phi(x), \mu(x), \nu(x), t) \label{evoc}
\eeqa
The inverse functional derivative 
$\left({\delta\over \delta \Phi(x)}\right)^{-1}$ is so defined:
\be
\left({\delta\over \delta \Phi(x)}\right)^{-1} 
\int {\cal D}k\; e^{-i\int k(y) \Phi(y) dy} 
= \int {\cal D}k\; {i\over k(x)}e^{-i\int k(y) \Phi(y) dy}  ~,
\ee
while the notation $\Delta [f(x)]$ stands for $f(x+\Delta x )- f(x)$. 
Performing an expansion in powers of $\hbar$ the classical limit may be 
obtained as in the previous sections.

\section{The Quantum Characteristic Function as a Generating Function}
In this section we discuss the connection between the 
probability representation described above both for quantum mechanics 
and quantum field theory and a slightly different point of view
developed in \cite{wet2}, where evolution equations are found for a 
suitably  defined euclidean partition function. 
There are two main ingredients in our
approach: one is is the probabilistic interpretation for the
distribution describing the observables, the other is the equivalence
between the description based on the MDF and the conventional
description based on the density matrix. The first aspect is guaranteed 
by the Glauber theorem which states that the Fourier transform of the
quantum characteristic function associated  to observables is a
probability distribution. The second aspect, namely the invertibility
of the MDF in terms of the density matrix, is achieved by introducing
configuration space variables which take value in an ensemble of
reference frames in phase space, each labelled by the two parameters, $\mu,
\nu$. 
Thus, the evolution equations which we have found (\eqn{evo}, \eqn{evon}, 
\eqn{evoc}), together with suitable initial conditions, completely 
characterize the state of the given quantum 
system. These equations assume a simpler form when their Fourier 
transform is performed. We have
\be
\chi(k,\mu,\nu,t)= \int dX e^{ikX} w(X,\mu,\nu,t)~,
\ee
with obvious generalizations to the $N$ dimensional case and to field 
theory.
For the one-dimensional quantum systems considered in section 1, 
the Fourier transform of Eq.\eqn{evo} yields an evolution equation for 
the quantum characteristic function itself:
\be
{\dot \chi}(k,\mu,\nu,t) = \left\{\frac{1}{m}
\mu {\del\over \del \nu}+{i\over 
\hbar} \left[ 
V\left({1 \over i k} {\del \over \del \mu} -{k\nu \hbar\over2} \right) 
- V\left({1 \over ik}  {\del \over \del \mu} 
+{k \nu \hbar\over2}\right)\right]\right\} \chi(k,\mu,\nu,t)~.
\label{evotr}
\ee
For the $N$-dimensional quantum systems considered in section 2, the 
Fourier transform of Eq.\eqn{evon} yields 
\beqa
{\dot \chi}(k_\sigma,\mu_\sigma,\nu_\sigma,t) &=& \Biggl\{
\frac{1}{m}\sum_{i=1}^N\mu_i {\del\over \del \nu_i}+{i\over \hbar} 
\left[ 
V\left(
{1\over i k_\sigma} {\del \over \del \mu_\sigma} 
-{\nu_\sigma k_\sigma \hbar\over2} 
\right)  \right. \nn\\
&-&\left. 
V\left( 
{1 \over i k_\sigma}  
{\del \over \del \mu_\sigma} 
+{\nu_\sigma k_\sigma\hbar\over 2} 
\right)
\right]
\Biggr\} \chi(k_\sigma,
\mu_\sigma,\nu_\sigma,t)~,
\label{evontr}
\eeqa
while Fourier transforming Eq.\eqn{evoc} we have, for quantum field 
theory,
\beqa
{\dot \chi}(k(x), \mu(x), \nu(x), t) &=& \left\{ \int d^d x~ 
\left[ 
\frac{1}{m}\mu(x) {\delta\over \delta \nu(x)} - 2i\hbar k(x)\nu(x)\Delta 
\left[ 
{1\over i k(x)}{\delta\over \delta \mu(x)}   
\right] 
\right] 
\right. \nn \\
&+& \left. i 
\left[
U\left[
{1\over i k(x)}  {\delta\over\delta \mu(x)} - {k(x) \nu(x)\hbar\over2} 
\right] \right.\right. \nn\\
&-&\left. \left. 
U\left[ 
{1\over i k(x)}  {\delta\over\delta \mu(x)} +{k(x) \nu(x)\hbar\over2} 
\right] \right] \right\}
\chi(k(x), \mu(x), \nu(x), t) \label{evoctr}~ .
\eeqa
Now the comparison with the results of \cite{wet2} may be easily understood.
Let us stick to quantum field theory for definiteness.
The quantum characteristic functional which is a generating functional 
for correlation functions of the $\Phi$ field 
coincides with the generating functional considered in
\cite{wet2}
\be
Z(\mu', \nu',t) =  \mbox{\rm Tr}  \left(
\hat\rho(t)\exp
\left\{ \int{ d^d x~ \mu'(x)\hat\phi(x)+ \nu'(x)\hat\pi(x) }
\right\}
\right)
~, \label{part}
\ee
after rescaling the parameters $\mu$ and $\nu$ to $\mu'=ik\mu$ and
$\nu'=ik\nu$ (of course, the same holds for quantum mechanics).
Consequently, the evolution equations for the characteristic functional
may be seen to be equal to those found in \cite{wet2} for the
generating functional \eqn{part} provided the parameters $\mu'$ and
$\nu'$ are rescaled as specified. 

Going back to the initial remark of this section, we may conclude that,
the quantum characteristic function and its evolution equation (or the
generating functional in \eqn{part}) are more interesting from an
operative point of view as they determine the correlation functions and
their time evolution. On the other hand the introduction of the MDF is
both relevant and necessary from a theoretical point of view. In facts
it allows a unified description of classical and quantum phenomena in
terms of probability distributions obeying different evolution
equations. Also it justifies the introduction of the $X$ variable, as a
variable taking values in an ensemble of reference frames \eqn{x}, in
view of the invertibility of the MDF in terms of the density matrix.
This seems to us the profound motivation for introducing such a
combination of phase space variables in the quantum characteristic
functional and in the generating functional \eqn{part} . We stress once
again that $X$ and its field analogue $\Phi$ are, for each couple
$(\mu,\nu)$, configuration space variables in the transformed reference
frame labelled by $(\mu,\nu)$.

\section{Conclusions}
In this letter we have presented an extension of the probabilistic
representation of quantum mechanics to quantum field theory. In this
framework classical and quantum phenomena, both statistically
described, only differ by the evolution equations of the distributions
of probabilities for the relevant observables. Quantum observables  are
described by a distribution of probability, the MDF, and the time
evolution by an integro-differential equation for the MDF. We recently
addressed the problem of finding the Green's function for the
time-evolution equation of the MDF \cite{noipra}. The problem was
solved for quadratic Hamiltonians, and a characterization of such a
propagator in terms of the time--dependent invariants of the system was
found. This propagator represents the transition {\it probability} of
the system from a quantum state to another. Thus, a generalization to
quantum field theory would be interesting in our opinion, and is
presently under consideration. 
Another promising application of the probabilistic point of view 
 is suggested in
\cite{wet2} where it is used to study the approach to equilibrium
of non equilibrium quantum field theories. 
An extension to relativistic quantum field theory would be also 
interesting, though it poses problems of interpretation which are not 
understood at the moment.

\end{document}